\documentclass[prd,12pt,a4paper,showpacs]{revtex4}
\usepackage{amssymb}
\usepackage{amsfonts}
\usepackage{amsbsy}

\def\dx{\delta\theta_{1(1)}}
\def\x{\theta}
\def\xo{\theta_1}
\def\xt{\theta_2}
\def\xn{\theta_{(0)}}
\def\xon{\theta_{1(0)}}
\def\xoo{\theta_{1(1)}}
\def\xtn{\theta_{2(0)}}
\def\xto{\theta_{2(1)}}
\def\b{\mathcal{B}}
\def\bo{\b_1}
\def\bt{\b_2}
\def\t{\Theta}
\def\to{\t_1}
\def\tt{\t_2}
\def\z{\beta}
\def\zo{\z_1}
\def\zt{\z_2}

\begin{document}

\title{Magnification relations for Kerr lensing \\ and testing Cosmic Censorship}

\author{M. C. Werner}
\email{mcw36@ast.cam.ac.uk}
\affiliation{Institute of Astronomy, University of Cambridge, Madingley Road, Cambridge CB3 0HA, United Kingdom}

\author{A. O. Petters}
\email{petters@math.duke.edu}
\affiliation{Departments of Mathematics and Physics, Duke University, 
Science Drive, Durham, NC 27708, United States of America}

\date{1 August 2007}

\begin{abstract}
A Kerr black hole with mass parameter $m$ and angular momentum parameter $a$ 
acting as a gravitational lens gives rise to two images in the weak field limit. 
We study the corresponding magnification relations, namely the 
signed and absolute magnification sums and the centroid up to post-Newtonian order. 
We show that there are post-Newtonian corrections to the total 
absolute magnification and centroid proportional to $a/m$, 
which is in contrast to the spherically symmetric case where such corrections vanish. Hence 
we also propose a new set of lensing observables for the two images 
involving these corrections, which should allow measuring $a/m$ with gravitational lensing. In fact, the 
resolution capabilities needed to observe this for the Galactic black hole 
should in principle be accessible to current and near-future instrumentation. Since $a/m >1$ 
indicates a naked singularity, a most interesting application would be a test of the Cosmic Censorship 
conjecture. The technique used to derive the image properties is based on the degeneracy of 
the Kerr lens and a suitably displaced Schwarzschild lens at post-Newtonian order. 
A simple physical explanation for this degeneracy is also given.

\textit{Version 2: to appear in Phys. Rev. D}, http://prd.aps.org/, \textit{Copyright (2007) by the American Physical Society.}
\end{abstract}

\pacs{04.25.Nx, 04.70.Bw, 95.30.Sf, 98.62.Sb}

\maketitle

\newpage

\section{Introduction}
The theory of gravitational lensing in the weak field limit has three 
physical inputs, namely perturbation theory of general relativity, 
geometric optics and the thin lens approximation \cite{schneider, petters}. 
Within this framework, a general formalism for lensing by spherically symmetric 
lenses up to post-post-Newtonian order in metric theories of gravity was 
recently developed by Keeton and Petters \cite{keeton1, keeton2}. 
Their appoach was extended to Kerr black holes by Sereno and 
de Luca \cite{sereno1} to study the effect of the angular momentum parameter 
on lensing properties.
 
We elaborate on this work in the present paper and determine the 
signed and total magnification sums of the images as well as the 
centroid for lensing by Kerr black holes in the weak field limit to 
post-Newtonian order. The post-Newtonian limit for rotating lenses was studied by 
Epstein and Shapiro \cite{epstein} and in more detail by 
Sereno \cite{sereno2}. Correction terms for the Kerr black hole up to post-post-Newtonian order 
were derived by Sereno and de Luca \cite{sereno1}, following earlier work by 
Bray \cite{bray}, from the equations of motion for null geodesics in Kerr geometry. 
Since we are presently interested in the post-Newtonian limit only, 
a considerably simpler method to derive image positions and magnifications can be 
applied, based on the analysis by Asada, Kasai, and Yamamoto \cite{asada}. 
This utilizes the degeneracy of rotating lenses and displaced 
non-rotating lenses at this order, shown to hold generally by 
Asada and Kasai \cite{asada2}. We provide a simple physical explanation for this 
degeneracy in the case of Kerr black holes in the weak field limit in Sec. II, 
then use the degeneracy explicitly to rederive image properties and 
find the new magnification relations in Sec. III. Based on this, we find new 
lensing observables for the two images involving post-Newtonian terms with $a/m$ in Sec. IV. 

This introduces a lensing technique to measure angular momentum parameters 
of rotating black holes, which could complement spectroscopic and photometric studies 
to this end. For instance, the supermassive black hole at the Galactic 
center (Sgr A*) shows flares in X-ray, infrared and radio bands with polarization 
and quasi-periodic ($\geq \mathrm{13 \pm 2 \ mins}$) substructure. Now this timescale 
appears to be associated with the innermost stable circular orbit, setting a lower limit 
of $a/m \geq 0.70 \pm 0.11$ (cf. \cite{trippe} and the discussion therein). Measurements 
of $a/m$ are important, in particular, since they indicate whether Kerr black holes 
have naked singularities, which is the case if $a/m>1$. The absence of naked 
singularities in nature is stipulated by the Cosmic Censorship conjecture 
which is evoked in the singularity theorems (see Penrose \cite{penrose} 
and references therein).

Hence the main application we have in mind is a lensing test 
of the Cosmic Censorship conjecture. Lensing properties of spherically symmetric 
static naked singularities were investigated by Virbhadra and Ellis \cite{virbhadra1}. 
More recently, Keeton and Petters \cite{keeton1} explored how lensing by the 
spherically symmetric Reissner-Nordstr\o m and Gibbons-Maeda-Garfinkle-Horowitz-Strominger 
black holes can be used to test the Cosmic Censorship conjecture. 
Here, however, we extend this work to the non-spherically symmetric but 
astrophysically more realistic Kerr black holes.

With regard to conventions, the metric 
signature $(-,+,+,+)$ is employed, and full units are used, where $G$ is the gravitational constant and
$c$ is speed of light, to facilitate observational applications. 
The mass parameter $m = G M_\bullet/c^2$ is the gravitational radius, 
where $M_\bullet$ is the 
physical mass of the black hole, and the angular momentum 
parameter $a=J/(M_\bullet c)$ is the specific angular momentum.
Greek indices denote spacetime coordinates and Latin indices spatial coordinates.

\section{Lensing framework}
\subsection{Post-Newtonian formalism}
We begin by reviewing the post-Newtonian formalism for a 
Schwarzschild lens with mass parameter $m$ in view of the later application to the 
Kerr black hole. 
Let $\boldsymbol{\t}=(\to,\tt), \ |\boldsymbol{\t}|=\t$ 
and $\boldsymbol{\b}=(\bo,\bt), \ |\boldsymbol{\b}|=\b$ 
be the Cartesian angular coordinates in the lens plane and source plane, 
respectively, whose coordinate axes are parallel and whose origins are on the 
optical axis. The deflection angle projected into the lens plane is denoted by 
$\boldsymbol{\hat{\alpha}}=(\hat{\alpha_1},\hat{\alpha_2}), \ 
|\boldsymbol{\hat{\alpha}}|=\hat{\alpha}$. Then straightforward plane geometry in the standard lensing framework yields the lens equation \cite{virbhadra2}
\begin{equation}
\tan \b=\tan \t -\frac{d_{LS}}{d_S}\left(\tan \t + \tan(\hat{\alpha} -\t)\right),
\label{lens}
\end{equation}
where $d_L,d_S,d_{LS}$ denotes the 
angular diameter distances from the observer to lens and source plane, 
and from the lens to the source plane, respectively. 
Up to the post-Newtonian limit, the angular coordinates can be expressed in terms of 
dimensionless coordinates 
$(\bo,\bt)=\theta_E(\zo,\zt),\ (\to,\tt)=\theta_E(\xo,\xt)$ 
such that
\begin{eqnarray*}
\b&=&\theta_E \z=\theta_E\left(\z_{(0)}+\z_{(1)}\epsilon+\mathcal{O}(\epsilon^2)\right) \\
\t&=&\theta_E \x=\theta_E\left(\x_{(0)}+\x_{(1)}\epsilon
+\mathcal{O}(\epsilon^2)\right),
\end{eqnarray*}
where the angular radius of the Einstein ring and the expansion parameter are
\begin{equation}
\theta^2_E=\frac{4md_{LS}}{d_Ld_S}, \ \ \epsilon=\frac{\theta_Ed_S}{4d_{LS}}.
\label{parameters}
\end{equation}
The impact paramter in the lens plane is $b=d_L \sin \t$ and the Schwarzschild deflection angle is
\[
\hat{\alpha}=4\frac{m}{b}+\frac{15\pi}{4}\frac{m^2}{b^2}+\mathcal{O}\left(\frac{m^3}{b^3}\right).
\]
Hence (\ref{lens}) can be recast thus
\begin{equation}
\z=\x-\frac{1}{\x}-\frac{15\pi}{16}\frac{\epsilon}{\x^2},
\label{lens2}
\end{equation}
to obtain the lens equation for the Schwarzschild black hole up to 
post-Newtonian order.

\subsection{Kerr lensing}

The line element of the Kerr metric $g^K_{\mu\nu}$ in the Boyer-Lindquist coordinates 
$\{t,r,\vartheta,\varphi\}$ denoted by $x^\mu_{BL}$ is \cite{chandra}
\begin{eqnarray}
ds^2&=&g^K_{\mu\nu}dx^\mu_{BL} dx^\nu_{BL}=-\left(1-\frac{2mr}{\rho^2} \right)c^2dt^2-\frac{4amr\sin^2\vartheta}{\rho^2}cdtd\varphi \nonumber \\
&+&\frac{\rho^2}{\Delta}dr^2+\rho^2d\vartheta^2+\frac{(r^2+a^2)^2-\Delta a^2\sin^2\vartheta}{\rho^2}\sin^2\vartheta d\varphi^2,
\label{kerr}
\end{eqnarray}
where $\rho^2=r^2+a^2\cos^2\vartheta$ and $\Delta=r^2-2mr+a^2$, 
which reduces to the Schwarzschild case if $a=0$. As mentioned above, the condition for a 
naked singularity is $a/m>1$ because then $\Delta >0$ and so no hypersurface $r=\mathrm{const.}$ 
can be null which in turn means that no event horizon exists. For $a\neq 0$, the 
degeneracy of the central caustic point of the Schwarzschild lens 
is lifted to give rise to a central caustic domain bounded by a 
distorted astroid \cite{rauch}. 
We are interested in the weak deflection limit and hence in the outer 
caustic domain where two images occur as in the Schwarzschild solution, 
albeit with modified positions and magnifications. In the standard gravitational lensing scenario, the null geodesics 
cross the equatorial plane $\vartheta=\pi/2$ at least once. In terms of constants of motion in Kerr geometry, 
we therefore restrict this discussion to null geodesics with Carter constant $Q\geq 0$ \cite[ p. 205]{oneill}.

The lens plane coordinates introduced in the previous section 
can now be conveniently oriented so that the $\tt$-axis
is along the projected angular momentum axis and forms a right-handed system 
together with $\to$-axis and the optical axis, with the observer at $d_L$, as the third one.
Now up to post-Newtonian order, Kerr lensing is equivalent to lensing by a 
Schwarzschild lens of the same mass but shifted to the position 
\cite{rauch, asada, asada2, sereno1}
\begin{equation}
\delta\boldsymbol{\t}=\theta_E(\delta\xo,0)=\theta_E(\dx \ \epsilon,0), \ \dx=\frac{a \sin \vartheta_O}{m},
\label{shift}
\end{equation}
where $\vartheta_O$ is the observer's polar angle position. 

We now show how this fact can be understood in a simple way in terms of the gravitomagnetic effect, and use 
it explicitly in the next section to find the corrected image positions and magnifications. 
For an extended discussion of the gravitomagnetic effect for the Kerr and more general rotating lenses, 
see Asada et al. \cite{asada, asada2}, Kopeikin et al. \cite{kopeikin} and Sereno \cite{sereno3}. In the weak field limit, 
the metric $g_{\mu\nu}$ can be understood as a formal perturbation 
$h_{\mu\nu}$ about the Minkowski metric $\eta_{\mu\nu}$ such that 
$g_{\mu\nu}=\eta_{\mu\nu}+h_{\mu\nu}, \ |h_{\mu\nu}|<<1$. Defining the trace-reversed perturbation 
$\overline{h}_{\mu\nu}=h_{\mu\nu}-g_{\mu\nu}h_{\alpha\beta}\eta^{\alpha\beta}/2$, 
Einstein's field equation may be written $\Box \overline{h}_{\mu\nu}
=-16\pi GT_{\mu\nu}/c^4$ in the de Donder gauge 
$\overline{h}^{\mu\nu}_{,\nu}=0$, where $T^{\mu\nu}$ is the energy momentum tensor. 
Now with a perfect, non-relativistic fluid, the components of its 
retarded solution give rise to a scalar field $U\equiv-\overline{h}_{00}c^2/4$, 
which is the Newtonian gravitational potential, and a vector potential with 
components $V^i\equiv\overline{h}_{0i}c^2$. 
Hence the perturbed metric line element is
\begin{equation}
ds^2=g_{\mu\nu}dx^\mu dx^\nu=-\left(1+ \frac{2U}{c^2} \right) c^2dt^2+\frac{2}{c}\sum_i V^i dx^i dt +\left( 1-\frac{2U}{c^2} \right)\sum_i(dx^i)^2 
\label{metric}
\end{equation}
where we work with spatially isotropic coordinates 
$\mathbf{x}=(x_1,x_2,x_3),|\mathbf{x}|=x$. Here $x_1$ is 
aligned with $\to$ and $x_3=0$ corresponds to the equatorial plane $\vartheta=\pi/2$. 
The gradient operator for this coordinate system is denoted by $\nabla$. Also, let $\mathbf{a}=a\mathbf{\hat{x}_3}$ where $\mathbf{\hat{x}_3}$ is the unit vector in the $x_3$ direction.

The equation of motion for null geodesics parametrized with $q$ and with 
unit ray 3-vector $\mathbf{k}$ can now be obtained from (\ref{metric}) 
using Fermat's principle, yielding a gravitoelectric and 
gravitomagnetic contribution (e.g., \cite{asada2}),
\begin{equation}
c^2\frac{d\mathbf{k}}{dq}=-2\nabla_\perp U+\mathbf{k}\times(\nabla \times \mathbf{V}),
\label{motion}
\end{equation}
where the operator $\nabla_\perp$ selects the component of the gradient perpendicular to the unit vector $\mathbf{k}$ such that, 
for any scalar field $\phi(\mathbf{x})$, $\nabla_\perp\phi\equiv\nabla\phi-(\nabla\phi\cdot\mathbf{k})\mathbf{k}=\mathbf{k}\times(\nabla\phi\times\mathbf{k})$. 
The forefactor of the gravitoelectric term is the well-known general 
relativistic correction of Newtonian light deflection. In this limit, the Kerr metric (\ref{kerr}) becomes
\begin{equation}
ds^2=-\left(1-\frac{2m}{x}\right)c^2dt^2-\frac{4am}{x^3}cdt(x_1dx_2-x_2dx_1)+\left(1+\frac{2m}{x}\right)\sum_i(dx^i)^2
\label{kerr2}
\end{equation}
whence one can read off $\mathbf{V}=-2GM_\bullet\mathbf{a}\times \mathbf{x}/x^3$ by 
comparison with (\ref{metric}).

We can now see that lensing due to a Kerr black hole at 
$\mathbf{x}=\mathbf{0}$ to post-Newtonian order is equivalent 
to a Schwarzschild lens displaced according to (\ref{shift}), that is, 
at $\delta \mathbf{x}=d_L\delta \boldsymbol{\t}$.
Since this Schwarzschild lens has zero vector potential 
and $U=-GM_\bullet/|\mathbf{x}-\delta\mathbf{x}|$, 
the right-hand side of (\ref{motion}) becomes, 
by Taylor expansion to post-Newtonian order,
\[
-2\nabla_\perp U(\mathbf{x}-\delta \mathbf{x})
= -2\nabla_\perp U(\mathbf{x})+2\nabla_\perp 
(\nabla U \cdot \delta \mathbf{x})
\]
because $\delta\mathbf{x}=\mathcal{O}(\epsilon)$ from (\ref{shift}), and the 
dot product is with respect to the Euclidean metric on the spatially isotropic coordinates. 
Furthermore, in the thin lens approximation, one may take $\mathbf{k}$ to be 
constant and perpendicular to the lens plane $L$ until $\mathbf{k}$ is 
changed by some $\delta\mathbf{k}$ upon crossing $L$. Since we 
consider the weak field limit, $|\delta\mathbf{k}|<<|\mathbf{k}|$ so the 
leading post-Newtonian term is
\[
2\nabla_\perp (\nabla U \cdot \delta \mathbf{x})=
2\mathbf{k}\times(\nabla(\nabla U \cdot \delta \mathbf{x})\times \mathbf{k})
= \mathbf{k}\times (\nabla \times (2\nabla U \cdot \delta \mathbf{x})\mathbf{k}),
\] 
which is indeed a gravitomagnetic term of the form occuring in (\ref{motion}). 
Hence the displaced Schwarzschild lens is equivalent to a point lens of mass $M_\bullet$ at $\mathbf{x}=\mathbf{0}$ with vector potential 
$2GM_\bullet a\sin \vartheta_O x_1\mathbf{k}/x^3$, using (\ref{shift}) and (\ref{parameters}). 
But $\mathbf{k}=(0,-\sin\vartheta_O,\cos\vartheta_O)$ by setup, so using the expression 
for $\mathbf{V}$ above we find that this vector potential component is precisely provided by a Kerr black hole situated at 
the origin with angular momentum parameter $a$, as required.

\section{Magnification relations}
\subsection{Image properties}

Following the discussion in the previous section, one can 
use the Schwarzschild lens equation to generate image 
properties of Kerr lensing up to post-Newtonian order in the weak field limit, 
where source and observer are in the asymptotically flat region with the source behind
the lens plane and close to the optical axis. 
Given the shift (\ref{shift}), we need to let $\xo\mapsto\xo-\delta\xo$ in the last two terms of (\ref{lens2}), 
which stem from the deflection angle of the lens model. Hence
\begin{eqnarray}
\zo&=&\xo-\frac{\xo-\delta\xo}{(\xo-\delta\xo)^2+\xt^2}-\frac{15\pi}{16}\frac{\xo-\delta\xo}{((\xo-\delta\xo)^2+\xt^2)^{3/2}}\ \epsilon+\mathcal{O}(\epsilon^2)\nonumber \\
\zt&=&\xt-\frac{\xt}{(\xo-\delta\xo)^2+\xt^2}-\frac{15\pi}{16}\frac{\xt}{((\xo-\delta\xo)^2+\xt^2)^{3/2}}\ \epsilon+\mathcal{O}(\epsilon^2)
\label{lens3}
\end{eqnarray}
is our ansatz for the Kerr lens equation. 
Notice that, at Newtonian order, (\ref{lens3}) reduces to the 
Schwarzschild lens equation (\ref{lens2}) for $\epsilon=0$ as expected, since 
$\delta\xo=\mathcal{O}(\epsilon)$  according to (\ref{shift}). 
The expansion of the image positions is
\begin{eqnarray}
\xo&=&\xon+\xoo \epsilon +\mathcal{O}(\epsilon^2) \nonumber \\
\xt&=&\xtn+\xto \epsilon +\mathcal{O}(\epsilon^2)
\label{position}
\end{eqnarray}
where $\xon,\xtn$ solve the lens equation at Newtonian order. 
This yields the two images of the 
well-known Schwarzschild case, one of positive and one of negative parity,
\begin{eqnarray}
\xon^\pm&=&\frac{\zo}{2}\left(1\pm\sqrt{1+\frac{4}{\z^2}}\right) \nonumber \\
\xtn^\pm&=&\frac{\zt}{2}\left(1\pm\sqrt{1+\frac{4}{\z^2}}\right)
\label{position1}
\end{eqnarray}
where $\xon^2+\xtn^2=\xn^2$.
Now at post-Newtonian order, the lens equation (\ref{lens3}) becomes
\begin{eqnarray*}
0&=&\xoo+\xon\left(\frac{A}{\xn^4}-\frac{15\pi}{16\xn^3}\right)-\frac{\xoo-\dx}{\xn^2}\\
0&=&\xto+\xtn\left(\frac{A}{\xn^4}-\frac{15\pi}{16\xn^3}\right)-\frac{\xto}{\xn^2}
\end{eqnarray*}
where $A=2(\xoo-\dx)\xon+2\xto\xtn$, and we recover the correction terms 
expected for 
rotating lenses \cite{sereno1},
\begin{eqnarray}
\xoo&=&\frac{15\pi\xon}{16(1+\xn^2)\xn}+\frac{(1-\xon^2+\xtn^2)\dx}{1-\xn^4}\nonumber\\
\xto&=&\frac{15\pi\xtn}{16(1+\xn^2)\xn}-\frac{2\xon\xtn\dx}{1-\xn^4}.
\label{position2}
\end{eqnarray}
Accordingly, the individual post-Newtonian corrections 
for the positive and the negative parity 
image are found by substituting $\xon^\pm, \xtn^\pm$ 
from (\ref{position1}) into (\ref{position2}). For a discussion and 
visualization of this shift, see Sereno \cite{sereno2}, especially his Figure 5. 

Since light rays are conserved in geometric optics, 
the signed image magnification $\mu$ is related to the Jacobian of the lens map 
\cite{schneider, petters},
\[
\frac{1}{\mu}=\det\left(\begin{array}{cc}
\frac{\partial \zo}{\partial \xo}&\frac{\partial \zo}{\partial \xt}\\
\frac{\partial \zt}{\partial \xo}&\frac{\partial \zt}{\partial \xt}
\end{array}\right).
\]
Recall also that the observable image flux $F_O$ and the 
flux of the unlensed source $F_S$ are related by $F_O=|\mu|F_S$.
Now evaluating the magnification yields
\begin{equation}
\mu=\frac{\xn^4}{\xn^4-1}-\left(\frac{15\pi\xn^3}{16(1+\xn^2)^3}
-\frac{4\xn^4\xon\dx}{(1-\xn^2)^2(1+\xn^2)^3}\right)\epsilon
+\mathcal{O}(\epsilon^2)
\label{mag}
\end{equation}
which, up to a sign, coincides with the findings in \cite{sereno1}. 
Again, this expression holds for both images.

\subsection{Magnification sums}

The magnification formula (\ref{mag}) can now be used together 
with (\ref{position1}) and (\ref{position2}) to write down a new expression for 
the individual magnifications of the positive and negative parity image, respectively, 
to post-Newtonian order,
\begin{eqnarray}
\mu^+ & = & \frac{(\z+\sqrt{4+\z^2})^4}{(\z+\sqrt{4+\z^2})^4-16}
- \frac{(2+\z^2+\z\sqrt{4+\z^2})(15\pi\z^3-64\zo\dx)}{8\z^3(\z+\sqrt{4+\z^2})^2(4+\z^2)^{3/2}}
\epsilon +   \mathcal{O}(\epsilon^2) \nonumber \\
\mu^- & = & \frac{(\z-\sqrt{4+\z^2})^4}{(\z-\sqrt{4+\z^2})^4-16}
- \frac{(2+\z^2-\z\sqrt{4+\z^2})(15\pi\z^3+64\zo\dx)}{8\z^3(\z-\sqrt{4+\z^2})^2(4+\z^2)^{3/2}}
\epsilon +\mathcal{O}(\epsilon^2).
\label{mag2}
\end{eqnarray}
Hence the sum of the signed magnifications can be evaluated and is of the simple form
\begin{equation}
\mu^+ +\mu^-=1-\frac{15\pi}{8(4+\z^2)^{3/2}}\epsilon+\mathcal{O}(\epsilon^2).
\label{mag3}
\end{equation}
The Schwarzschild lens obeys a well-known magnification invariant 
(e.g., \cite[ p. 191]{petters}) in the standard lensing framework, that is,
at lowest order. Since the signed magnification sum (\ref{mag3}) for the Kerr lens 
does not depend on the specific angular momentum $a$, 
it is identical to the Schwarzschild lens result to post-Newtonian order 
(cf. Eq. (54) of \cite{keeton2}). The Kerr lens has thus the same 
deviation from the magnification invariant as the Schwarzschild lens at $\mathcal{O}(\epsilon)$.

Now taking into account the image parities, the absolute magnifications 
are $|\mu^+|=\mu^+$ and $|\mu^-|=-\mu^-$.  Hence, the total absolute magnification
is 
\begin{equation}
\mu_{\rm tot} = 
|\mu^+|+|\mu^-|=\frac{2+\z^2}{\z\sqrt{4+\z^2}}+\frac{8\zo}{\z^3(4+\z^2)^{3/2}}\frac{a\sin\vartheta_O}{m}\epsilon+\mathcal{O}(\epsilon^2)
\label{mag4}
\end{equation}
using (\ref{parameters}). 
The term $\mathcal{O}(\epsilon)$ vanishes for $a=0$ or an observer on the 
rotational axis of the Kerr black hole, that is $\vartheta_O=0$, as expected for 
circularly symmetric lenses \cite{keeton1}.

\subsection{Centroid}

We can also define the centroid of the magnification thus,
\[
\mathbf{\Theta^{\rm Cent}}
=\theta_E\frac{\boldsymbol{\x}^+|\mu^+|+\boldsymbol{\x}^-|\mu^-|}{|\mu^+|+|\mu^-|}.
\]
Given (\ref{position1}), (\ref{position2}) and (\ref{mag2}), 
a new expression for the centroid vector of Kerr images to post-Newtonian order 
can now be obtained, and its components turn out to be
\begin{eqnarray}
\Theta^{\rm Cent}_1 &=&\theta_E\left[\frac{(3+\z^2)\zo}{2+\z^2}
+\frac{(\zo^2-\zt^2-2)}{(2+\z^2)^2}\frac{a\sin\vartheta_O}{m}\epsilon
+\mathcal{O}(\epsilon^2)\right] \nonumber \\
\Theta^{\rm Cent}_2 &=&\theta_E\left[\frac{(3+\z^2)\zt}{2+\z^2}
+\frac{2\zo\zt}{(2+\z^2)^2}\frac{a\sin\vartheta_O}{m}\epsilon
+\mathcal{O}(\epsilon^2)\right].
\label{centroid}
\end{eqnarray}
Again, in the circularly symmetric case $a=0$ or $\vartheta_O=0$ we can take $\zt=0$ without 
loss of generality, to recover the result by Keeton and Petters \cite{keeton1}.

\section{Applications}

\subsection{Breaking the degeneracy}

Asada and Kasai \cite{asada} found that, at post-Newtonian order, 
rotating and non-rotating dark lenses 
cannot be distinguished on account of the degeneracy discussed in Sec. II, 
that is, by observing the images alone. 
This problem can be circumvented if the location of the black hole is 
established independently, for instance by 
observing the center of the accretion disk surrounding the Kerr black hole. 
To see this, recall from Sec. II that the 
Kerr lens $K$ is equivalent to a displaced Schwarzschild lens up to post-Newtonian 
order in the weak field limit 
where observer and source are assumed to be in the asymptotically flat region 
of the Kerr black hole. 
Hence a plane $P$ containing the observer, the source and the notional shifted 
Schwarzschild lens will also 
contain the two images of the Kerr lens as for a standard Schwarzschild lens. 
Projected into the plane of the sky, the 
source, the notional shifted Schwarzschild lens and the two images will be collinear 
but not typically with $K$ since $K\notin P$ in general. 
Hence, if the position of $K$ can be found independently, the projected distance of 
$K$ from the line joining the two 
images is observable and the degeneracy is broken in the generic case. 
However, note for completeness that there 
are very special cases in which the degeneracy cannot be broken in this way: 
consider a source such that $\bt=0$ exactly, 
so $\xtn=0$ and hence $\xto=0$ from (\ref{position2}). 
In this case, $K \in P$ and so the projected source, shifted Schwarzschild lens position, and the
two images will all be collinear with $K$ at the origin. Therefore, the degeneracy is not broken.

But assuming the Kerr/Schwarzschild degeneracy is broken successfully, 
we still need to be able to measure image positions 
in the $(\to,\tt)$ coordinate system in order to apply the formalism. 
Hence the direction of the Kerr black hole's 
spin axis projected into the lens plane must also be known and, 
moreover, the observer's $\vartheta_O$ coordinate. In principle, this 
could be inferred from observations of the jet associated with the black hole 
accretion disk because of the frame-dragging effect on the magnetohydrodynamics 
of the jet (e.g., \cite{narayan}). Furthermore, time-dependent measurements of the polarization 
of black hole flare emission could constrain the direction of the spin axis. In the case of Sgr A*, for 
instance, this seems to indicate that the black hole spin axis is essentially 
aligned with the Galaxy's \cite{trippe}.

\subsection{Measuring the angular momentum parameter}

In order to determine whether Kerr lensing could be used to 
measure $a/m$ and test the Cosmic Censorship conjecture, 
we first of all need to assemble a suitable set of observables. This 
set is in turn dependent on resolution capability: 
If the two Kerr images can in fact be resolved, then image positions and fluxes will be 
observable individually. If, however, they cannot be resolved, 
the total flux and the magnification centroid may be taken as observables. 
We shall discuss the former case first.

If the two images can be resolved, two vectorial image positions and two fluxes are available, 
giving six equations altogether. Using (\ref{position}-\ref{position2}) and (\ref{mag3}-\ref{mag4}), 
we propose the following combinations as convenient observables,
\begin{eqnarray}
\to^+ +\to^-&=&\frac{\bo}{\sqrt{4\theta_E^2+\b^2}}\left(\sqrt{4\theta_E^2+\b^2}
-\frac{15\pi\theta_E}{16}\epsilon\right)
+\frac{a\sin\vartheta_O\theta_E}{m}\epsilon +\mathcal{O}(\epsilon^2) 
\label{eq:newobservables-eqs1}\\
\tt^+ +\tt^-&=&\frac{\bt}{\sqrt{4\theta_E^2+\b^2}}\left(\sqrt{4\theta_E^2+\b^2}
-\frac{15\pi\theta_E}{16}\epsilon\right)+\mathcal{O}(\epsilon^2)\\
\to^+ -\to^-&=&\frac{\bo}{\b}\left(\sqrt{4\theta_E^2
+\b^2}+\frac{15\pi\theta_E}{16}\epsilon\right)
-\frac{\b^4\theta_E+4\bt^2\theta_E^3}{\b^3\sqrt{4\theta_E^2+\b^2}}\frac{a\sin\vartheta_O}{m}
\epsilon +\mathcal{O}(\epsilon^2) \\
\tt^+ -\tt^-&=&\frac{\bt}{\b}\left(\sqrt{4\theta_E^2+\b^2}+\frac{15\pi\theta_E}{16}
\epsilon\right)
+\frac{4\bo\bt\theta_E^3}{\b^3\sqrt{4\theta_E^2+\b^2}}\frac{a\sin\vartheta_O}{m}\epsilon 
+\mathcal{O}(\epsilon^2) \\
F_O^+ + F_O^-&=&F_S\left(\frac{2\theta_E^2+\b^2}{\b\sqrt{4\theta_E^2+\b^2}}+\frac{8\bo\theta_E^5}{\b^3(4\theta_E^2+\b^2)^{3/2}}\frac{a\sin\vartheta_O}{m}\epsilon\right)+\mathcal{O}(\epsilon^2)\\
F_O^+ - F_O^-&=&F_S\left(1-\frac{15\pi\theta_E^3}{8(4\theta_E^2+\b^2)^{3/2}}\epsilon\right)
+\mathcal{O}(\epsilon^2)
\label{eq:newobservables-eqs6}
\end{eqnarray}
which reduce to the formulae of Keeton and 
Petters \cite{keeton2} for $a=0$ or $\sin \vartheta_O=0$, as required. 
These six equations could then be solved for the six occuring variables $\bo,\bt,F_S,\theta_E,\epsilon,a\sin\vartheta_O/m$. Assuming that the lensed 
source orbits the black hole such that $d_{LS}<<d_L$ and that an independent estimate for $m$ is available, then $d_{LS}$ 
and $d_L \approx d_S$ can also be found from $\theta_E$ and $\epsilon$ using (\ref{parameters}). 

It would therefore be possible to infer $a/m$ from the post-Newtonian lensing corrections. Moreover, these 
corrections should in principle be observable with near-future instrumentation as discussed by 
Keeton and Petters \cite{keeton2}. In the case of Sgr A*, they
found an estimate for the angular Einstein radius to be of order
$\theta_E = 0.022 (d_{LS}/ 1 {\rm pc})^{1/2}\ \mathrm{arcsec}$ and  
perturbation parameter to be of order 
$\epsilon=2.1 \times 10^{-4} \times (d_{LS}/ 1 {\rm pc})^{1/2}$.  
Since the image separations will be of order $\theta_E$, these images
can in principle be resolved with current technology (e.g., the CHARA interferometer
and radio interferometry can resolve $10^{-3} \ \mathrm{arcsec}$ separations \cite{brumm}). 
Furthermore, currently known positional uncertainties in observed radio images  
are of order $10^{-6} \ \mathrm{arcsec}$
(e.g., \cite{trotter}).   
These are indeed much smaller than the current resolution capabilities. 
In addition,
the statistical prospects for observing lensed stars around 
Sgr A* are discussed in a forthcoming paper by 
Congdon et al. \cite{congdon} who conclude that the disk component of the 
Milky Way contributes more than the bulge, 
and find that the expected number of lenses reaches unity for a 
detection limit of $K\sim \mathrm{18.5 \ mag}$. 
  
In the case when the two images cannot be resolved, then only the total flux (22) 
and the centroid (\ref{centroid}) are available. In the 
forseeable future, this situation applies to lensing by 
extragalactic supermassive black holes, and Congdon et al. \cite{congdon} estimate that 
typically $\sim 100$ lensed stars can be expected. 
But given that we only have three equations for six variables in this case, 
a determination of $a/m$ does not seem possible. However, 
this situation may improve if additional information, for instance on $m$, $d_L$ and $F_S$, 
becomes available.

Finally, we should stress again that the success of this method is conditional, in both cases, 
upon breaking the Kerr/Schwarzschild degeneracy and establishing the $\to,\tt,\vartheta_O$ 
coordinates, as discussed in the previous section. Further data, for example the time delay 
between images of a variable source (cf. \cite{keeton2}) or images of multiple sources, may also be helpful for 
breaking degeneracies. Since equations (\ref{eq:newobservables-eqs1} - \ref{eq:newobservables-eqs6}) 
fully determine the six occurring variables, we have not considered these ramifications here. 
Nonetheless, our analysis shows that lensing measurements of $a/m$ for supermassive black holes, 
and hence lensing tests of Cosmic Censorship, have potential.

\section{conclusion}

We considered gravitational lensing in the weak field limit of a 
Kerr black hole of mass parameter $m$ and angular momentum parameter $a$ and derived the 
magnification relations for the two ensuing images up to post-Newtonian order. 
The image properties used were rederived with a simple perturbation analysis 
based on the degeneracy of a Kerr lens and a Schwarzschild lens shifted 
by (\ref{shift}). Whereas the signed magnification sum (\ref{mag3}) 
turned out to be identical to the Schwarzschild case, 
the absolute magnification sum (\ref{mag4}) and centroid (\ref{centroid})
show a term proportional to $a/m$ at post-Newtonian order. 
This is in contrast to circularly symmetric lenses where these terms have 
been shown to vanish precisely. In discussing observational implications, 
we provided a new set of six lensing observables 
(\ref{eq:newobservables-eqs1} - \ref{eq:newobservables-eqs6}) 
for the case that the two images can be resolved. These are matched with
six lensing variables including $a\sin\vartheta_O/m$. 
In the case of lensing by the Galactic black hole, 
the two images should be resolvable by current and near-future instrumentation, 
so that measurements of the angular momentum parameter should be feasible. 
Since $a/m>1$ for naked singularities, 
this provides a possible test of the Cosmic Censorship conjecture using 
gravitational lensing. Additional study of this issue is definitely warranted.

\section*{Acknowledgments}
MCW would like to thank the Department of Mathematics, Duke University, for their hospitality and gratefully acknowledges funding by the STFC (formerly PPARC), United Kingdom.
AOP strongly acknowledges the research funding support of NSF Grants
DMS-0302812, AST-0434277, and AST-0433809.

\end{document}